 \documentclass[10pt,preprint]{aastex}  

\def\gapprox{\lower.4ex\hbox{$\;\buildrel >\over{\scriptstyle\sim}\;$}}
\def\lapprox{\lower.4ex\hbox{$\;\buildrel <\over{\scriptstyle\sim}\;$}}

\def\ref#1{\par\noindent\hangindent1cm {#1}}

\shortauthors{ASCHWANDEN 2010}
\shorttitle{Self-Organized Criticality Model}

\begin{document}

\title{		A Universal Scaling Law for the Fractal Energy Dissipation
		Domain in Self-Organized Criticality Systems }

\author{        Markus J. Aschwanden}

\affil{         Lockheed Martin Advanced Technology Center,
                Solar \& Astrophysics Laboratory,
                Org. ADBS, Bldg.252,
                3251 Hanover St.,
                Palo Alto, CA 94304, USA;
                e-mail: aschwanden@lmsal.com}

\begin{abstract}
Nonlinear dissipative systems in the state of self-organized criticality
release energy sporadically in avalanches of all sizes, such as in 
earthquakes, auroral substorms, solar and stellar flares, soft gamma-ray
repeaters, and pulsar glitches. The statistical
occurrence frequency distributions of event energies $E$ generally exhibit
a powerlaw-like function $N(E)\propto E^{-\alpha_E}$ with a powerlaw slope
of $\alpha_E \approx 1.5$. The powerlaw slope $\alpha_E$ of energies can
be related to the fractal dimension $D$ of the spatial energy dissipation
domain by $D=3/\alpha_E$, which predicts a powerlaw slope $\alpha_E=1.5$
for area-rupturing or area-spreading processes with $D=2$. 
For solar and stellar flares,
2-D area-spreading dissipation domains are naturally provided in current 
sheets or separatrix surfaces in a magnetic reconnection region. Thus,  
this universal scaling law provides a useful new diagnostic on the topology 
of the spatial energy dissipation domain in geophysical and astrophysical 
observations.
\end{abstract}


\section{       INTRODUCTION  			}

The statistics of catastrophic events in geophysics (e.g., earthquakes,
landslides, forest fires) as well as in astrophysics (e.g., auroral
substorms, solar and stellar flares, pulsar glitches) is generally
quantified in a log-number versus log-size histogram, which often exhibits
a powerlaw-like function, also called occurrence frequency distribution. 
The most widely known example is the distribution of earthquakes magnitudes, 
which has a powerlaw slope of $\alpha \approx 2.0$ for the differential
frequency distribution (Turcotte 1999), the so-called Gutenberg-Richter (1954) 
law.  Bak, Tang, and Wiesenfeld (1987, 1988) introduced the theoretical concept
of self-organized criticality (SOC), which has been initially applied to
sandpile avalanches at a critical angle of repose, but has been
generalized to nonlinear dissipative systems that are driven 
in a critical state. Comprehensive reviews on this subject can be found 
for applications in geophysics (Turcotte 1999), solar physics 
(Charbonneau et al.~2001), and astrophysics (Aschwanden 2010). 

Hallmarks of SOC systems are the scale-free powerlaw distributions of various 
event parameters, such as the peak energy dissipation rate $P$, 
the total energy $E$, or the time duration $T$ of events. 
While the powerlaw shape of the distribution
function can be explained by the statistics of nonlinear processes that have
an exponential growth phase and saturate after a random time interval 
(e.g., Willis and Yule 1922; Fermi 1949; Rosner and Vaiana 1978; 
Aschwanden et al.~1998), no general theoretical model has been developed that
predicts the numerical value of the powerlaw slope of SOC parameter
distributions. Simple analytical models that characterize the nonlinear
growth phase with an exponential growth time $\tau_G$ and the random
distribution of risetimes with an average value of $t_S$, predict a
powerlaw slope of $\alpha_P = 1 + t_S/\tau_G$ for the peak dissipation rate
(e.g., Rosner and Vaiana 1978; Aschwanden et al.~1998), but observations 
of the time scales $t_S$ and $\tau_G$ require time profiles with an 
exponentially-growing risetime, which can reliably be determined in simple 
single-peak events only, while larger events often show complex multi-peak 
time profiles. Moreover, an estimate of the powerlaw slope $\alpha_E$ of
energies requires an additional relationship between the peak energy
dissipation rate $P$ and the total energy $E$. An alternative theoretical 
explanation for a slope $\alpha_E=3/2$ was put forward by a dimensional 
argument (Litvinenko 1998), which can be derived from the definition of
the kinetic energy $E \propto M (L/T)^2$ of convective flows with mass $M$, 
length scale $L$, and time scale $T$. Note, that this dimensional argument
predicts a unique fixed value $\alpha_E$ and presupposes kinetic motion for 
an energy dissipation process. In this Letter we propose a more
general concept where the powerlaw slope of the energy distribution
depends on the fractal geometry of the energy dissipation domain of 
SOC events.

\section{	THEORETICAL CONCEPT 		}

We aim for a universal statistical model of nonlinear energy dissipation 
processes in systems that operate in a state of self-organized criticality.
We start with the spatial structure of nonlinear energy dissipation 
domains. Let us consider a system with volume $V_0$ that is subject
to continuous energy input. Examples are sandpiles (on which sand is
steadily dropped), the Earth crust (which is sheared by
tectonic plates), or the solar corona (to which magnetic energy emerges,
generated by the solar dynamo from the interior). Once such
a system is in a state of self-organized criticality, the steadily
stored energy is sporadically dissipated in small or large catastrophic
events, such as in form of sandpile avalanches, earthquakes, or solar
flares. In the statistical average we may assume that the energy input
is homogeneously distributed throughout the volume $V_0$ that is 
susceptible to energy dissipation, while energy dissipation processes
occur randomly at different places and times, with a time-averaged
energy dissipation rate that matches the average energy input rate.
However, the size or magnitude of individual dissipation events is
unpredictable and affects an arbitrary subvolume $V_E \le V_0$. The
subvolume affected in an energy dissipation event depends very much
on the propagation characteristics of the instability that triggers
the catastropic energy dissipation event. A lightening may propagate
mostly along a 1-dimensional (1-D) path, snow avalanches may spread
over a 2-dimensional (2-D) area, earthquakes may spread along 2-D
tectonic plates, solar flares may propagate along 2-D separatrix
surfaces in magnetic reconnection regions, or a fire may spread in
a 3-dimensional (3-D) volume wherever flammable material is available. 
We may characterize the volume $V_E$ affected in a nonlinear energy 
dissipation domain with a fractal dimension $D$ as function of the 
length scale $L$,
$$
	V_E(L) \propto L^{D} \ ,
	\eqno(1)
$$
where $D$ denotes the volumetric fractal dimension $0 < D \le 3$, 
e.g., as defined by the Hausdorff dimension. 
Assuming a constant energy
density $dE/dV = const$  in the time average, the total released energy $E$ 
is than proportional to the fractal volume,
$$
	E = V_E(L) \left({dE \over dV}\right) \propto L^{D} \ .
	\eqno(2)
$$
Energy dissipation events can have all possible spatial sizes $L$ up to 
the maximum length scale $L_0=V_0^{-3}$ of the system. Considering 
the probability of fragmentation into larger or smaller scales,
the statistical ratio between different size scales must be reciprocal 
to the volume in the time average, if energy balance is required between
the time-averaged input rate and the time-averaged dissipation rate. 
This simple volume fragmentation rule directly yields the probability 
or occurrence frequency distribution $N(E)$ of released energies $E$
(using Eq.~2),
$$
	N(E) \propto V^{-1} \propto L^{-3} \propto E^{-3/D} \ .
	\eqno(3)
$$
We see that the assumption of a fractal dimension automatically leads
to a powerlaw distribution of energies, at least over some range of sizes (i.e.,
the {\sl inertial range}), which is one of the hallmarks of systems with
self-organized criticality. 
This simple statistical argument let us predict the following energy
distributions for the three integer fractal dimensions $D=1, 2, 3$,
(but the fractal dimension $D$ can also take any non-integer value in the 
range of $0 < D \le 3$), 
$$
	\begin{array}{ll}
	N(E) \propto E^{-3.0} & {\rm for} \ D=1 \\
	N(E) \propto E^{-1.5} & {\rm for} \ D=2 \\
	N(E) \propto E^{-1.0} & {\rm for} \ D=3 \\
	\end{array}
	\eqno(4)
$$
For example, dissipation events with a 1-D propagation path (e.g., 
lightenings) are expected to produce frequency distributions of 
$N(E) \propto E^{-3}$, while events with 2-D propagation paths
(e.g., tectonic plates, magnetic separator surfaces) are expected
to produce distributions of $N(E) \propto E^{-1.5}$. In the following we
concentrate mostly on the latter case with area-like $(D=2)$
propagation characteristics, which is known in geophysics as the type 
of {\sl area rupture process} (e.g., Turcotte 1999). 
The spatial fragmentation law is visualized
in Fig.~1 for a cartesian geometry of a cube with volume $V_0$, which is
divided into length scales of $L_i=2^{-i}=1, 1/2, 1/4, ...$, where the area
rupture in each subcube scales as $A_i=L_i^2=2^{-2i}=1, 1/4, 1/16, ...$, 
and the number of subcubes is $n_i=(2^i)^3= 1, 8, 64, ...$. If the
released energy is proportional to the area, $E_i \propto A_i \propto L_i^2$,
the released energy per cube is $E_i = 1, 1/4, 1/16, ...$, and 
the resulting occurrence frequency distribution has then the powerlaw slope 
$\alpha = - \ln(n_{i+1}/n_i) / \ln(E_{i+1}/E_i) =-1.5$.

Frequency distributions of energies, or proxy parameters of the energy,
are measured for a substantial number of phenomena associated with
nonlinear dissipative systems in the state of self-organized criticality,
which generally display a powerlaw-like distribution function over some
inertial range,
$$
	N(E) dE \propto E^{-\alpha_E} dE \ .
	\eqno(5)
$$
If we set the observed distribution function (Eq.~5) equal to the
theoretical model (Eq.~3), we obtain from the observable powerlaw slope
$\alpha_E$ a diagnostic of the fractal dimension $D$ of the propagation
characteristics of the energy release process,
$$
	D = {3 \over \alpha_E} \ .
	\eqno(6)
$$  
For instance, if a powerlaw distribution of $\alpha_E=1$ is observed,
we expect a space-filling or Euclidian propagation characteristics
($D=3$), such as a diffusion-like or fractional Brownian motion 
process (which, however, is not a SOC process). Powerlaws with 
$\alpha_E=1.5$ are expected to have area-like rupturing or 
spreading characteristics ($D=2$), while powerlaws with 
$\alpha_E=3$ are expected to have linear propagation 
characteristics ($D=1$). Thus, the observed frequency distributions
convey fundamental information on the fractal geometry of the
energy dissipation domain.
 
\section{	DISCUSSION 			}

In the following we compile measurements of observed occurrence
frequency distributions of energies in SOC phenomena, or proxy
parameters of energies. We quote only measurements that refer to the
total energy of a SOC event, such as the thermal, nonthermal, magnetic,
or kinetic energy, or the time-integrated radiative output in time
profiles (i.e., the time-integrated counts or fluence) in astrophysical 
observations. Table 1 lists the SOC phenomena, the observed powerlaw
slopes $\alpha_E$ of energy parameters, and the fractal dimension
$D=3/\alpha_E$ inferred with Eq.~(6). We briefly discuss the results
in the following.

Solar flares observed in hard X-rays are believed to provide the most direct
measure of dissipated energy. An example of a frequency distribution of
nonthermal energy in electrons at energies of $>25$ keV is shown in
Fig.~2, which shows a powerlaw with a slope of $\alpha_E=1.53\pm0.02$
extending over an energy range of four orders of magnitude, i.e.,
$E=10^{28}-10^{32}$ erg (Crosby, Aschwanden, and Dennis 1993). 
This powerlaw slope implies a fractal dimension of $D\approx 1.9-2.0$,
which is perfectly consistent with a fractal 2-D energy dissipation domain,
such as current sheets or separatrix surfaces in magnetic reconnection
regions (Fig.~3, top). 
An early cellular automaton model was able to reproduce this value
consistently for different system sizes (Lu et al.~1993). Other measurements
of hard X-ray energies yield similar values in the range of $\alpha_E
\approx 1.4-1.7$ (Lee et al.~1993; Bromund et al.~1995; Perez-Enriquez and
Miroshnichenko 1999; Georgoulis et al.~2001; Christe et al.~2008), which
are approximately consistent with a fractal dimension of $D\approx 2$. 

Solar flares observed in soft X-rays provide a measure of the thermal
energy, which are also found to be in the range of $\alpha_E=1.44-1.6$
(Drake et al.~1971; Shimizu 1995), except for one measurement with a
somewhat higher value of $\alpha_E=1.88$ (Veronig et al.~2002a,b), which 
may be affected by background-subtraction issues. The general coincidence
between energy distributions in soft and hard X-rays can be taken as 
evidence for the Neupert effect, which implies that the thermal flare plasma
is evaporated from the chromosphere by heating through precipitating nonthermal
particles. Thus, although the spatial geometry of soft X-ray radiating
postflare loops is 1-D, their energy distribution reflects the 2-D geometry
of current sheets where the primary energy dissipation took place.

For solar nanoflares observed in EUV we see two trends: (1) Powerlaw slopes
of energy distributions inferred from single EUV temperature filters have 
values in the range of $\alpha_E \approx 1.8-2.6$ (Krucker and Benz 1998;
Parnell and Jupp 2000; Aschwanden et al.~2000; Benz and Krucker 2002),
which suffer from the bias of underestimating the thermal energy due to the
lack of high-temperature filters; (2) Powerlaw slopes inferred from 
combined EUV and soft X-ray temperature filters yield a value of
$\alpha_E=1.54\pm0.03$ (Aschwanden and Parnell 2002), which is again
consistent with a fractal dimension of $D=3/1.54 \approx 2$ as derived
from flare energies observed in soft and hard X-rays.

Solar energetic particle (SEP) events were reported to have 
flatter powerlaw distributions, i.e., $\alpha_E \approx 1.2-1.4$ (Gabriel
and Feynman 1996), which are preferentially associated with the largest
flares, and thus their statistics is subject to a selection effect that
is not representative for all flares.

Stellar flares were found to have a large range of powerlaw slopes,
in the range of $\alpha_E \approx 1.3-2.4$ (Audard et al. 2000; 
Robinson et al.~1999), but each star provides only very few events 
per observing run (typically $\approx 5-15$), and thus the large spread
can be explained by small-number statistics. 

Soft-gamma repeaters were reported to have burst fluence distributions with
powerlaw slopes of $\alpha_E \approx 1.4-1.7$ (Gogus et al.~1999, 2000; Chang
et al.~1996) and implies fractal geometries of $D\approx 1.8-2.1$, 
which may reflect area-like fractures in neutron star crusts strained 
by evolving magnetic stresses (Thompson and Duncan 1995). 

In magnetospheric physics, the energy of geomagnetic substorms 
was estimated from the area sizes of auroras (Fig.~3, middle)
and the dissipated power, which were
found to have frequency distributions with powerlaw slopes of
$\alpha_E \approx 1.0-1.2$ (Lui et al.~2000) and imply a fractal
dimension of $D =3 / \alpha_E \approx 3$. This suggests an Euclidean or 
volume-filling spreading of energy dissipation, which may
reflect the 3-D geometry of the plasmoid that forms during the
expansion phase in the geotail (Baumjohann and Treuman 1996).

In geophysics, earthquakes were found to have powerlaw distributions of
$\alpha_E \approx 2.0$ (Turcotte 1999), which implies a fractal dimension
of $D \approx 1.5$, possibly reflecting a mixture of 1-D earthquake fault
ruptures (Fig.~3, bottom)
and 2-D ruptures or tectonic plate spreading. Landslides were found in a
range of $\alpha_E\approx 1.7-3.3$, which also implies a mixed dimensionality
of $D \approx 0.9-1.8$, ranging from 1-D avalanches guided along local
valleys to 2-D avalanches that spread over flat hill sides. Forest fires
were reported to have powerlaw slopes of $\alpha_E\approx 1.3-1.5$
(Turcotte 1999), which is perfectly consistent with 2-D area-spreading
in extended forest regions. Even city sizes were found to follow a
powerlaw slope of $\alpha_E\approx 1.4$ (Zanette 2007), which corresponds
to the urban sprawl over 2-D areas. 

Besides energy distributions, also the distributions of spatial scales
of SOC events can shed some light into the fractal geometry of the energy
dissipation regions. According to Eq.~3 we expect a fragmentation scaling of
$N(L) \propto L^{-3}$. Such a spatial fragmentation scaling law is indeed
confirmed for the sizes or Saturn rings, i.e., 
$N(L) \propto L^{-3}$ (Zebker et al. 1985; French and Nicholson 2000);
for the sizes of asteroids, i.e.,
$N(L) \propto L^{-2.3}... L^{-4}$ (Ivezic et al.~2001); 
and for sizes of lunar craters, i.e., a cumulative distribution
$N^{cum}(>L) \propto L^{-2}$ (Cross 1966), which corresponds to the 
differential distribution $N(L) \propto L^{-3}$.

\section{	CONCLUSIONS 			}

In conclusion, our universal model for fractal energy dissipation domains
in systems with self-organized criticality predicts (i) the powerlaw shape
of occurrence frequency distributions of energies, as well as (ii) a
relationship of $\alpha_E=3/D$ between the the powerlaw slope $\alpha_E$
and the fractal dimension $D$ of the energy dissipation domain.
Most observed occurrence frequency distributions of SOC events
exhibit a powerlaw-like function with a slope of $\alpha_E \approx 1.5$, 
which is predicted for instabilities that have 2-D area-like propagation 
characteristics, such as tectonic plate ruptures in earthquakes or
current sheets in magnetic reconnection regions. This new scaling law
could be corroborated by investigating the geometric morphology of energy
dissipation events and their statistical frequency distributions in
observations with spatial imaging capabilities, while it provides 
geometric predictions for non-imaging astrophysical observations. 

\acknowledgements {\sl Acknowledgements:} 
This work is partially supported by NASA contract
NAS5-98033 of the RHESSI mission through University of California,
Berkeley (subcontract SA2241-26308PG). 


\section*{REFERENCES}

\def\ref#1{\par\noindent\hangindent1cm {#1}}

\ref{Aschwanden, M.J., Dennis, B.R., and Benz, A.O. 1998, ApJ 497, 972.}

\ref{Aschwanden, M.J., Tarbell, T., Nightingale, R., Schrijver, C.J.,
        Title, A., Kankelborg, C.C., Martens, P.C.H., and Warren, H.P.
        2000, ApJ 535, 1047.}

\ref{Aschwanden, M.J. and Parnell, C.E. 2002, ApJ 572, 1048.}

\ref{Aschwanden, M.J., 2010, {\sl Self-Organized Criticality in Astrophysics.
	Statistics of Nonlinear Processes in the Universe},
	Springer-PRAXIS: New York (in press).}

\ref{Audard, M., G\"udel, M., Drake, J.J., and Kashyap, V.L. 2000,
        ApJ 541, 396.}

\ref{Bak, P., Tang, C., and Wiesenfeld, K. 1987, Phys. Rev. Lett. 59/27, 381.}

\ref{Bak, P., Tang, C., and Wiesenfeld, K. 1988, Phys. Rev. A 38/1, 364.}

\ref{Baumjohann, W. and Treuman, R.A. 1996, {\sl Basic Space Plasma Physics},
        Imperial College Press: London.}

\ref{Benz, A.O. and Krucker, S. 2002, ApJ 568, 413.}

\ref{Bromund, K.R., McTiernan, J.M., and Kane, S.R. 1995, ApJ 455, 733.}

\ref{Chang, H.K., Chen, K., Fenimore, E.E., and Ho, C. 1996,
        AIP Conf. Proc. 384, 921.}

\ref{Charbonneau, P., McIntosh, S.W., Liu, H.L., and Bogdan, T.J. 2001,
        Solar Phys. 203, 321.}

\ref{Christe, S., Hannah, I.G., Krucker, S., McTiernan, J., and Lin, R.P.
        2008, ApJ 677, 1385.}

\ref{Crosby, N.B., Aschwanden, M.J., and Dennis, B.R. 1993, 
	Solar Phys. 143, 275.}

\ref{Crosby, N.B., Meredith, N.P., Coates, A.J., and Iles, R.H.A. 2005,
        Nonlinear Processes in Geophysics 12, 993.}

\ref{Cross, C.A. 1966, MNRAS 134, 245.}

\ref{Drake, J.F. 1971, Solar Phys. 16, 152.}

\ref{Fermi, E. 1949, Phys. Rev. Lett. 75, 1169.}

\ref{French, R.G. and Nicholson, P.D. 2000, Icarus 145, 502.}

\ref{Gabriel, S.B. and Feynman, J. 1996, Solar Phys. 165, 337.}

\ref{Galsgaard, K., Priest, E.R., and Nordlund, A. 2000, Solar Phys. 193, 1.}

\ref{Georgoulis, M.K., Vilmer,N., and Crosby,N.B. 2001, A\&A 367, 326.}

\ref{Gogus, E., Woods, P.M., Kouveliotou, C., van Paradijs, J.,
        Briggs, M.S., Duncan, R.C., and Thompson, C. 1999, ApJ 526, L93.}

\ref{Gogus, E., Woods, P.M., Kouveliotou, C., and van Paradijs, J. 2000,
        ApJ 532, L121.}

\ref{Gutenberg, B. and Richter, C.F. 1954,
        {\sl Seismicity of the Earth and Associated Phenomena},
        Princeton University Press, Princeton, NJ, p.310 (2nd ed.).}

\ref{Ivezic, Z., Tabachnik, S., Rafikov, R., Lupton, R.H., Quinn, T.,
        Hammergren, M., Eyer, L., Chu, J., Armstrong, J.C., Fan, X.,
        Finlator, K., Geballe, T.R., Gunn, J.E., Hennessy, G.S., Knapp, G.R.,
        et al. (SDSS Collaboration) 2001, AJ 122, 2749.}

\ref{Krucker, S. and Benz, A.O. 1998, ApJ 501, L213.}

\ref{Lee, T.T., Petrosian, V., and McTiernan, J.M. 1993, ApJ 412, 401.}

\ref{Litvinenko, Y.E. 1998, Solar Phys. 180, 393.}

\ref{Lu, E.T., Hamilton, R.J., McTiernan, J.M., and Bromund, K.R. 1993,
        ApJ 412, 841.}

\ref{Lui, A.T.Y., Chapman, S.C., Liou,K., Newell, P.T., Meng, C.I.,
        Brittnacher, M., and Parks, G.K. 2000, GRL 27/7, 911.}

\ref{Parnell,C.E. and Jupp,P.E. 2000, ApJ 529, 554.}

\ref{Perez-Enriquez, R. and Miroshnichenko, L.I. 1999, Solar Phys. 188, 169.}

\ref{Robinson, R.D., Carpenter, K.G., and Percival, J.W. 1999, ApJ 516, 916.}

\ref{Rosner, R., and Vaiana, G.S. 1978, ApJ 222, 1104.}

\ref{Shimizu, T. 1995, Publ. Astron. Soc. Japan 47, 251.}

\ref{Thompson, C. and Duncan, R.C. 1995, MNRAS, 275, 255.}

\ref{Turcotte, D.L. 1999, Rep. Prog. Phys. 62, 1377.}

\ref{Veronig, A., Temmer, M.,Hanslmeier, A., Otruba, W., and Messerotti, M.
        2002a, A\&A 382, 1070.}

\ref{Veronig, A., Temmer, M., and Hanslmeier, A. 2002b,
        Hvar Observatory Bulletin 26/1, 7.}

\ref{Willis, J.C. and Yule, G.U. 1922, Nature 109, 177.}

\ref{Zanette, D.H. 2007, in  "The Dynamics of Complex Urban Systems. 
	An Interdisciplinary Approach",  (eds. S.  Albeverio, D. Andrey, P. 
	Giordano, and A. Vancheri, eds. (Springer, Berlin, 2007).}

\ref{Zebker, H.A., Maroufm, E.A., and Tyler, G.L. 1985, Ikarus 64, 531.}

\clearpage

\begin{deluxetable}{llll}
\tabletypesize{\footnotesize}
\tablecaption{Observed occurrence frequency distributions of energies 
(or fluences) in systems with self-organized criticality.
{\sl References}:
$^1$) Crosby et al.~(1993),
$^2$) Lu et al.~(1993), 
$^3$) Lee et al.~(1993),	
$^4$) Bromund et al.~(1995),
$^5$) Perez-Enriquez and Miroshnichenko (1999), 
$^6$) Georgoulis et al.~(2001), 
$^7$) Christe et al.~(2008),
$^8$) Drake et al.~(1971),
$^9$) Shimizu (1995), 
$^{10}$) Veronig et al.~(2002a, 2002b),
$^{11}$) Krucker and Benz~(1998), 
$^{12}$) Parnell and Jupp~(2000), 
$^{13}$) Aschwanden et al.~(2000), 
$^{14}$) Benz and Krucker~(2002), 
$^{15}$) Aschwanden and Parnell~(2002), 
$^{16}$) Gabriel and Feynman~(1996), 
$^{17}$) Audard et al.~(2000), 
$^{18}$) Robinson et al.~(1999),
$^{19}$) Gogus et al.~(1999, 2000),
$^{20}$) Chang et al.~(1996), 
$^{21}$) Lui et al.~(2000),
$^{22}$) Crosby et al.~(2005), 
$^{23}$) Zanette (2007),
$^{24}$) Turcotte (1999).} 
\tablewidth{0pt}
\tablehead{
\colhead{SOC Phenomenon}&
\colhead{Powerlaw slope $\alpha_E$}&
\colhead{Fractal dimension $D$}&
\colhead{Reference}}
\startdata
Solar flare, hard X-rays&1.53$\pm$0.03  & 1.9-2.0	&1)	\\
			&1.51		& 2.0		&2)	\\
			&1.62		& 1.9		&3)	\\
			&1.74		& 1.7		&4)	\\
			&1.39		& 2.1		&5)	\\
			&1.39		& 2.1		&6)	\\
			&1.7		& 1.8		&7)	\\
Solar flares, soft X-rays &1.44		& 2.1		&8)	\\
			&1.5-1.6	& 1.9-2.0	&9)	\\
			&1.88		& 1.6		&10)	\\
Solar flares, EUV	&2.3-2.6	& 1.2-1.3	&11)	\\
			&2.0-2.6	& 1.2-1.5	&12)	\\
			&1.79$\pm$0.08	& 1.6-1.8	&13)	\\
			&2.0-2.6	& 1.2-1.5	&14)	\\
			&1.54$\pm$0.03	& 1.9-2.0	&15)	\\
Solar energetic particles&1.2-1.4	& 2.1-2.5	&16)	\\
Stellar flares		&1.3-2.4	& 1.2-2.3       &17)	\\
			&2.0		& 1.5		&18)	\\
Soft gamma repeaters	&1.66		& 1.8		&19)	\\
              		&1.4-1.7	& 1.8-2.1       &20)	\\
Auroral blobs		&1.0-1.2	& 2.5-3.0	&21)	\\
Outer radiation belt	&1.5-2.1	& 1.4-2.0	&22)	\\
City sizes              &1.4		& 2.1           &23)	\\
Forest fires		&1.3-1.5	& 2.0-2.3	&24)	\\
Earthquakes		&2.0		& 1.5		&24)	\\
Landslides		&1.7-3.3        & 0.9-1.8	&24)	\\
\enddata
\end{deluxetable}


\begin{figure}
\plotone{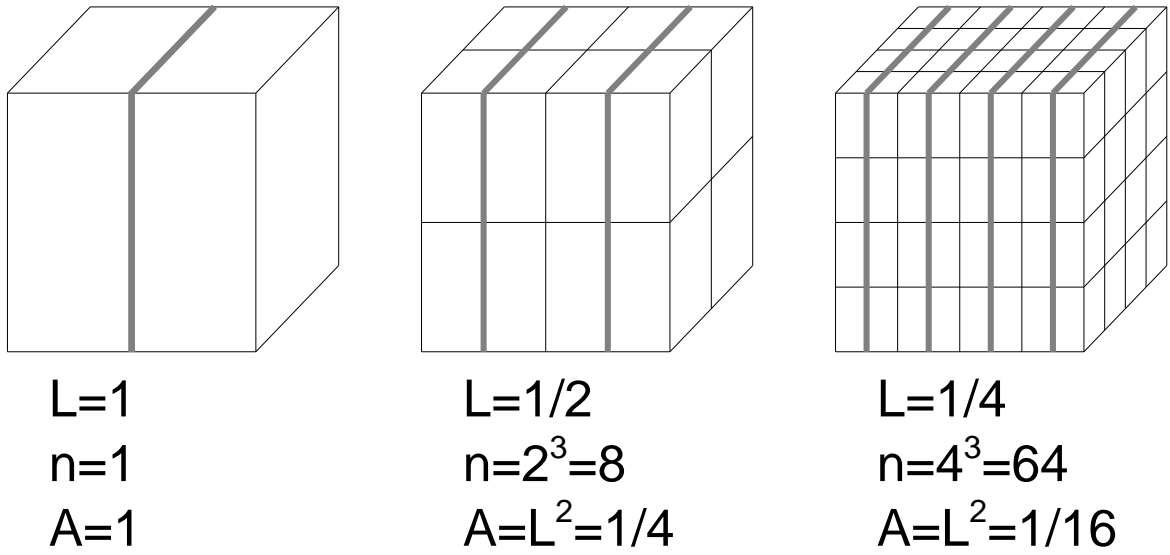}
\caption{Schematic diagram of an area-fracturing process (grey lines) 
in a 3-D cube.
The length scales of subcubes decrease by a factor 2 in each step
($L_i=2^{-i}, i=0,1,2$), while the number of cubes increases by 
$n_i=(2^i)^3$, and the fractured area in each subcube scales as
$A_i=L_i^2$. Energy dissipation is expected to occur only in the
areas $A_i$, which yields a statistical frequency distribution $N(E)
\propto E^{-1.5}$ of energies $E$, if $E \propto A$.}
\end{figure}

\begin{figure}
\plotone{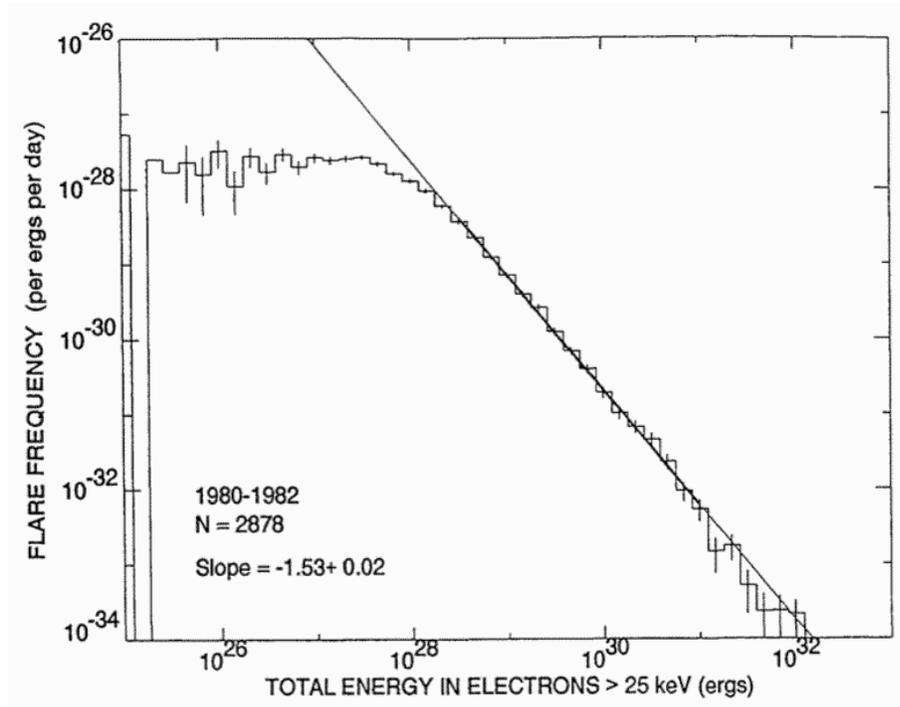}
\caption{The frequency distribution of the total energy in electrons above
25 keV calculated by integrating the thick-target energy rate over the
flare duration, from a set of 2878 flare events observed with the
{\sl Hard X-Ray Burst Spectrometer (HXRBS)} onboard the {\sl Solar
Maximum Mission (SMM)}. The powerlaw fit yields a slope of
$\alpha_E=1.53\pm 0.02$, (Crosby et al.~1993).}
\end{figure}

\begin{figure}
\plotone{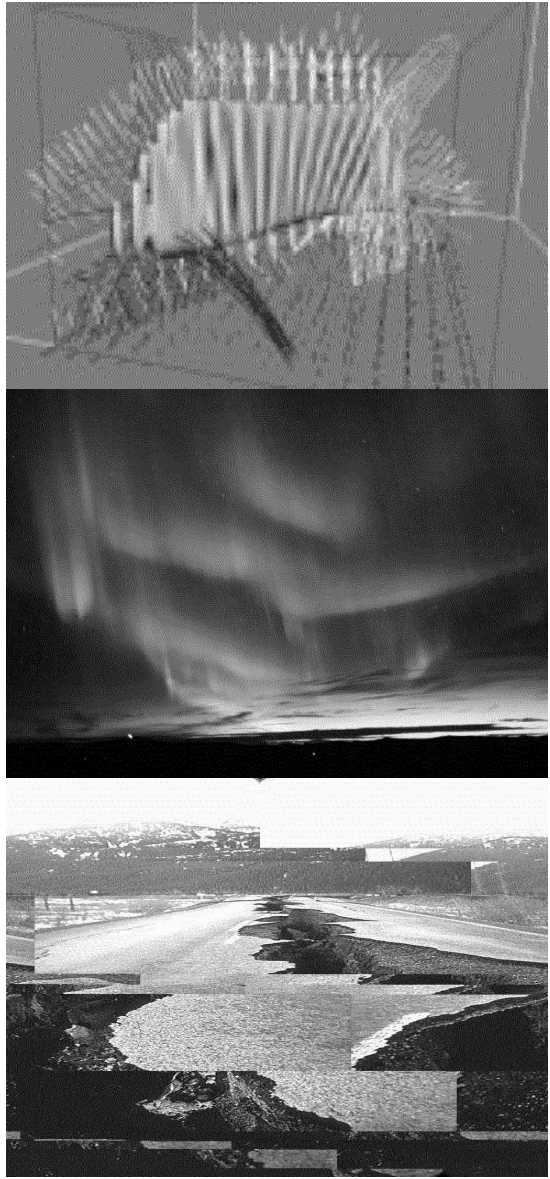}
\caption{Examples of 2-D area-spreading energy dissipation events:
{\sl Top:} 2-D iso-surface of current accumulation in a magnetic 
reconnection process, calculated with a 3-D MHD code (Galsgaard et al.~2000);
{\sl Middle:} 2-D sheet structure of aurora borealis seen in Alaska 
(credit: {\sf www.geology.com/nasa/}); 
{\sl Bottom:} Rupture on the surface of a road during an
earthquake in Alaska 1964 (credit: {\sf www.sciencecourseware.org}).}
\end{figure}

\end{document}